\documentclass[twocolumn,amsmath,amssymb,nofootinbib,prl,superscriptaddress,eqnumsec]{revtex4-1}
\pdfoutput=1

\usepackage{graphicx}
\usepackage{dcolumn}
\usepackage{bm,psfrag}
\usepackage{mathrsfs}

\usepackage[retainorgcmds]{IEEEtrantools}

\newcommand{\be}{\begin{equation}}
\newcommand{\ee}{\end{equation}}
\newcommand{\bea}{\begin{eqnarray}}
\newcommand{\eea}{\end{eqnarray}}
\newcommand{\ba}{\begin{eqnarray}}
\newcommand{\ea}{\end{eqnarray}}

\newcommand{\beq}{\begin{equation}}
\newcommand{\eeq}{\end{equation}}
\newcommand{\beqa}{\begin{eqnarray}}
\newcommand{\eeqa}{\end{eqnarray}}
\newcommand{\beqar}{\begin{eqnarray*}}
\newcommand{\eeqar}{\end{eqnarray*}}

\def\t6 {T_\mt{D6}}

\newcommand{\mt}[1]{\textrm{\tiny #1}}

\def\cale         {{\cal E}}

\def\ee           {{\rm e}}

\def\Re           {{\rm Re\hskip0.1em}}

\def\sqr#1#2{{\vcenter{\vbox{\hrule height.#2pt
 \hbox{\vrule width.#2pt height#1pt \kern#1pt
 \vrule width.#2pt}\hrule height.#2pt}}}}

\def\ee{\cale}

\def\aa1{\phi}
\def\cc1{\psi}

\usepackage[bookmarks=false]{hyperref} 
\hypersetup{pdfstartview=FitH,pdfhighlight=/O,colorlinks=false}

\begin{document}

\title{Nonlinear Evolution and Final Fate of Charged Anti--de Sitter Black Hole
  Superradiant Instability}

\author{Pablo Bosch}
\email{pbosch@perimeterinstitute.ca}
\affiliation{Department of Physics \& Astronomy and Guelph-Waterloo Physics Institute\\
University of Waterloo, Waterloo, Ontario N2L 3G1, Canada}
\affiliation{Perimeter Institute for Theoretical Physics, Waterloo, Ontario N2L 2Y5,
Canada}
\author{Stephen R. Green}
\email{sgreen@perimeterinstitute.ca}
\author{Luis Lehner}
\email{llehner@perimeterinstitute.ca}
\affiliation{Perimeter Institute for Theoretical Physics, Waterloo, Ontario N2L 2Y5,
Canada}

\begin{abstract}
  We describe the full nonlinear development of the superradiant
  instability for a charged massless scalar field, coupled to general
  relativity and electromagnetism, in the vicinity of a
  Reissner--Nordstr\"om-AdS black hole. The presence of the negative
  cosmological constant provides a natural context for considering
  perfectly reflecting boundary conditions and studying the dynamics
  as the scalar field interacts repeatedly with the black hole.  At
  early times, small superradiant perturbations grow as expected from
  linearized studies.  Backreaction then causes the black hole to lose
  charge and mass until the perturbation becomes nonsuperradiant, with
  the final state described by a stable hairy black hole.  For large
  gauge coupling, the instability extracts a large amount of charge
  per unit mass, resulting in greater entropy increase. We discuss the
  implications of the observed behavior for the general problem of
  superradiance in black hole spacetimes.
\end{abstract}

\maketitle

{\bf Introduction.}---A bosonic field can extract energy from a
rotating black hole (BH) through superradiant
scattering~\cite{Zeldovich:1972,1973ZhETF..64...48S}, resulting in an
increase in field amplitude. If, in addition, the field is reflected
by a potential barrier sufficiently far away, then this amplification
process repeats, resulting in exponential growth. This is known as the
superradiant instability or ``black hole
bomb''~\cite{Press:1972zz}. Since in global anti--de Sitter (AdS)
spacetime, naturally reflecting boundary conditions at $\mathscr{I}$
can be defined, asymptotically AdS black holes with ergoregions are
subject to the superradiant
instability~\cite{Hawking:1999dp,Cardoso:2004hs,Green:2015kur}.

A similar process occurs for Reissner-Nordstr\"om (RN) BHs~\cite{Christodoulou:1972kt,Denardo:1973,Bekenstein:1973mi,Hawking:1999dp},
with the charge playing the role of the angular momentum. There, a charged scalar field mode with time dependence
$\psi \sim e^{-i\omega t}$ is superradiantly amplified if
$\omega r_H < qQ$ ($r_H$ is the BH outer horizon radius,
$Q$ is the BH charge, and $q$ is the gauge coupling of the
scalar field). In the RN-AdS
case~\cite{Hawking:1999dp,2011PhRvD..83f4020U}, the reflecting
boundary implies that there is a minimum mode frequency, so the
instability sets in when $Qq>\frac{3r_H}{L}$, where $L$ is the AdS scale
(in the limit of small $r_H/L$).

When perturbations are small, the description above is
valid and a linearized analysis is suitable; several studies have
determined the quasinormal mode spectra of, e.g.,
Kerr-AdS~\cite{Cardoso:2004hs,Cardoso:2013pza} and
RN-AdS~\cite{2011PhRvD..83f4020U}. As the perturbation grows, however,
the backreaction on the spacetime becomes significant, and this
description breaks down.

Less is known about the final state of the instability, although it is,
in general, expected to be a ``hairy'' BH. In the RN-AdS case, static
BHs surrounded by a scalar field condensate have been constructed, and
have been conjectured to be the end point of the
instability~\cite{Basu:2010uz,Dias:2011tj}. In the Kerr-AdS case,
rotating BHs with a single helical Killing vector field have been
constructed~\cite{Dias:2015rxy}, but these BHs are unstable themselves
and are, therefore, not plausible end points. The final state might be
a hairy BH without any symmetries, or the instability may lead to a
violation of cosmic censorship~\cite{Niehoff:2015oga}.  Further
complication arises from the fact that gravitational interactions can
result in significant nonlinear mode coupling in confined geometries
such as AdS, where dissipation is
low~\cite{Bizon:2011gg,Balasubramanian:2014cja}.

It is of wide interest to have a more complete picture of the dynamics
and end point of the superradiant instability. In astrophysics, the
instability is used to constrain dark matter models and may lead to
observable gravitational wave
emissions~\cite{2011PhRvD..83d4026A,Yoshino:2012kn,2012PhRvL.109m1102P}. In
holography (AdS/CFT~\cite{Maldacena:1997re} and
Kerr/CFT~\cite{Guica:2008mu}) superradiance manifests within the
CFT~\cite{Dias:2007nj,Murata:2008xr,Bredberg:2009pv,Compere:2012jk,Cardoso:2013pza},
and the nonlinear evolution plays a role in determining the final
thermal state. Finally, questions of BH instabilities are of
theoretical interest in classical general
relativity~\cite{Dafermos:2010hb,Dafermos:2014cua,Green:2015kur}.

Fully analyzing the superradiant instability is complicated by several
factors: large differences in scale between the BH and perturbation,
long instability time scales, fully nonlinear equations (Einstein and
other fields), and an intrinsically $(3+1)$-dimensional problem. Thus,
nonlinear simulations, while the obvious approach, are
challenging~\cite{Lehner:2001wq,2015arXiv150206853C}; however,
see~\cite{2014PhRvD..89f1503E}. In the charged case, however, the
instability is present even in spherical symmetry, and the instability
time scale is shorter. By studying this case, one could hope to draw
general conclusions that could be applied more broadly.

In this Letter we study the nonlinear evolution of the superradiant
instability of RN-AdS BHs in spherical symmetry. We verify the initial
growth rates predicted from the linear theory, and we confirm
expectations that the final state is a hairy BH\footnote{These results
  are consistent with very recent results
  of~\cite{Sanchis-Gual:2015lje} for the case of RN surrounded by an
  artificial mirror (which appeared while we were completing our
  work). Working in AdS provides a natural setting free of ambiguities
  and potential sources of constraint violations. }.  At intermediate
times, we track the dynamics of individual modes as charge and mass
are nonlinearly extracted from the BH, and we arrive at an intuitive
picture of the behavior. In particular, within this picture, the
behavior characterized as a ``bosenova''
of~\cite{Sanchis-Gual:2015lje} for large values of $q$ can be easily
understood.

{\bf Model.}---We follow the conventions
of~\cite{Gubser:2008px}, and we work in $d=4$ dimensions. The
Lagrangian density is
\begin{equation}
  \label{eq:lagrangian}
  16\pi G_N\mathcal{L} = R + \frac{6}{L^2} - \frac{1}{4}F_{ab}F^{ab} - |D_a \psi|^2,
\end{equation}
where $D_a\equiv\nabla_a - i q A_a$ is the gauge covariant
derivative. This gives rise to the Einstein equation
\begin{equation}\label{eq:einstein}
  G_{ab} - \frac{3}{L^2}g_{ab} = 8\pi T_{ab}^\psi + 8\pi T_{ab}^{\text{EM}},
\end{equation}
where the stress-energy tensors are
\begin{IEEEeqnarray}{rCl}
  8\pi T_{ab}^{\text{EM}} & = & \frac{1}{2}\left( g^{cd} F_{ac}F_{bd} - \frac{1}{4}g_{ab}F_{cd}F^{cd}\right),\\
  8\pi T_{ab}^\psi & = & \frac{1}{2}\left[(D_a\psi)^\ast(D_b\psi) + \text{c.c.}\right] 
  -\: \frac{1}{2} g_{ab}|D_c\psi|^2.
\end{IEEEeqnarray}
The Maxwell and scalar field equations are
\begin{IEEEeqnarray}{rCl}
  \label{eq:maxwell}\nabla^b\left(\nabla_b A_a - \nabla_a A_b\right) &=& i q \psi^\ast D_a\psi - i q \psi(D_a\psi)^\ast,\\
  \label{eq:scalar}D^aD_a\psi &=&0.
\end{IEEEeqnarray}

The RN-AdS BH solves the field equations, with metric
$ds^2 = -f dt^2 + f^{-1} dr^2 + r^2 d\Omega_2^2$, where
$f = 1 - \frac{2M}{r} + \frac{Q^2}{4r^2} + \frac{r^2}{L^2}$;
Maxwell field $A_\mu dx^\mu = \left( \frac{Q}{r} - \frac{Q}{r_H} \right) dt$;
and $\psi=0.$ The $-\frac{Q}{r_H}$ term in the Maxwell field is a gauge choice
to have the field vanish at the horizon.

{\bf Numerical Method.}---Our simulations follow the general approach
of~\cite{Chesler:2013lia}. We adopt ingoing Eddington-Finkelstein
coordinates and spherical symmetry\footnote{This differs from the
  analysis of~\cite{Murata:2010dx}, which imposed planar symmetry and
  studied the nonlinear evolution of a holographic superconductor.}
so that the metric takes the form
\begin{equation}\label{eq:metricform}
  ds^2 = - A(v,r) dv^2 + 2 dv dr + \Sigma(v,r)^2 d\Omega_2^2.
\end{equation}
For the Maxwell field, we work in a gauge where 
\begin{equation}\label{eq:Maxwellform}
  A_\mu dx^\mu = W(v,r) dv,
\end{equation}
and we require $\psi=\psi(v,r)$.  With these choices, the equations of
motion \eqref{eq:einstein}, \eqref{eq:maxwell}, and \eqref{eq:scalar}
take the form
\begin{IEEEeqnarray}{rCl}
  \IEEEeqnarraymulticol{3}{l}{\text{\underline{Einstein:}}}\nonumber\\
  \label{eq:Einstein1}0&=& \Sigma (d_+\Sigma)' + (d_+\Sigma) \Sigma' - \frac{3}{2L^2}\Sigma^2 - \frac{1}{2} + \frac{1}{8} \Sigma^2 W', \\
  \label{eq:Einstein2}0&=& A'' - \frac{4}{\Sigma^2}(d_+\Sigma)\Sigma' + \frac{2}{\Sigma^2} + (\psi')^\ast d_+\psi \nonumber\\
  && +\: (d_+\psi)^\ast \psi' - (W')^2 + iqW\left[ \psi^\ast\psi' - (\psi')^\ast\psi \right],\\
  \label{eq:Einstein3}0&=& d_+d_+\Sigma -\frac{1}{2} A' d_+\Sigma +\frac{1}{2}\Sigma|d_+\psi|^2 + \frac{1}{2}q^2W^2\Sigma|\psi|^2 \nonumber\\ 
  && +\: \frac{1}{2}iqW\Sigma\left[\psi^\ast d_+\psi - \psi(d_+\psi)^\ast\right],\\
  \label{eq:Einstein4}0&=& \Sigma'' + \frac{1}{2}\Sigma |\psi'|^2,\\
  \IEEEeqnarraymulticol{3}{l}{\text{\underline{Maxwell:}}}\nonumber\\
  \label{eq:Maxwell1}0&=& (d_+W)' - \frac{1}{2}A'W' + 2 \frac{d_+\Sigma}{\Sigma}W' - 2q^2W|\psi|^2 \nonumber\\
  &&+\:iq\left[\psi^\ast d_+\psi - \psi (d_+\psi)^\ast\right],\\
  \label{eq:Maxwell2}0&=& W'' +\frac{2}{\Sigma} \Sigma' W' +iq \left[ \psi^\ast\psi' - \psi (\psi')^\ast \right],\\
  \IEEEeqnarraymulticol{3}{l}{\text{\underline{Scalar:}}}\nonumber\\
  \label{eq:Scalar1}0&=& 2(d_+\psi)' + 2\frac{\Sigma'}{\Sigma}d_+\psi + 2 \frac{d_+\Sigma}{\Sigma}\psi'  - iq\psi W' \nonumber\\
  && -\: 2iq\frac{\Sigma'}{\Sigma}W\psi - 2iqW\psi',
\end{IEEEeqnarray}
where we denote $f'\equiv\partial_rf$,
and the derivative along the outgoing null direction,
$d_+f\equiv \partial_vf + \frac{1}{2}A\partial_rf$.

The equations of motion are solved imposing reflecting boundary
conditions at $r\to\infty$. These take the form
\begin{IEEEeqnarray}{rCl}
  \label{eq:Aasymp}A &=& \frac{r^2}{L^2} + \lambda r + \left(1 + \frac{L^2\lambda^2}{4} - L^2\dot{\lambda}\right)- \frac{2M}{r} \nonumber\\ 
  && +\: \left(L^2\lambda M + \frac{Q^2}{4}\right)\frac{1}{r^2} +  O(r^{-3}),\\
  \Sigma &=& r + L^2\lambda/2 + O(r^{-5}),\\
  W &=& {\nu} + Q/r + O(r^{-2}),\\
  \label{eq:psiasymp}\psi &=& \varphi_3/r^3 + O(r^{-4}).
\end{IEEEeqnarray}
The constants $M$ and $Q$ represent the ADM mass and charge,
respectively; these are prescribed as boundary data. The functions
$\lambda(v)$ and $\nu(v)$ represent the remaining gauge freedom after
putting the metric and Maxwell fields into the forms
\eqref{eq:metricform}--\eqref{eq:Maxwellform}. We make the further
gauge choice that $\lambda=\nu=0$. Finally the function $\varphi_3(v)$
is an unknown function that is determined by the solution.

The procedure to integrate the equations is as follows: On an initial
time slice $v=v_0$ we prescribe the function $\psi(v_0,r)$. We then
integrate, radially inwards in $r$ [and subject to the asymptotic conditions
\eqref{eq:Aasymp}--\eqref{eq:psiasymp}], eqs.~\eqref{eq:Einstein1},
\eqref{eq:Einstein2}, \eqref{eq:Einstein4}, \eqref{eq:Maxwell2}, and
\eqref{eq:Scalar1}, to obtain $d_+\Sigma$, $A$, $\Sigma$, $W$, and
$d_+\psi$, respectively, at time $v=v_0$. From $d_+\psi$, $\psi$, and
$A$, we obtain $\partial_v\psi$ at time $v=v_0$. By integrating in
time, we obtain $\psi$ at the next time step. The procedure may then
be iterated. Equations \eqref{eq:Einstein3} and \eqref{eq:Maxwell1}
are redundant, and we use these as independent residuals to test our
code.

We use finite differences, using a mixed second and fourth order
radial and fourth order in time Runge-Kutta method. The spatial domain
extends from an inner radius $r_0$---several grid points within the
BH---to infinity. In fact, we compactify this domain by introducing a
new spatial coordinate $\rho = 1/r$.  This gives rise to a compact
domain $0\le \rho \le 1/r_0$, which we discretize in a uniform grid.

For initial data, we take the scalar field to be compactly supported,
with
$\psi = (r^{-1} - r_1^{-1})^3 (r^{-1} - r_2^{-1})^3/r^2 [\kappa_1 +
\kappa_2 \sin(10/r)]$
if $r\in[r_1,r_2]$ (and zero otherwise).  For production runs we
typically use $\{\kappa_1, \kappa_2\} = 10^{-4}$, so the scalar field
is initially negligible compared to the BH, and $\{r_1,r_2\}=\{2M,3M\}$. 
We checked by varying
$\{\kappa_1,\kappa_2\}$ that different initial data do not affect the
features of the final solution, provided the amplitude is small. We
set $L=1$, $M=0.1$ and $Q=-0.18115$; this
corresponds to a BH small compared to the AdS scale ($r_H=0.138$),
with charge $63.9\%$ of the critical value.  Excellent accuracy is
obtained with grid sizes of $N= 1600n + 1$ points (with $n=1$ for low
 $q$ and $n=2,3$ for higher ones), and we have thoroughly
tested our implementation (see the Supplemental Material).

{\bf Results.}---Having fixed $M$ and $Q$, we varied the gauge
coupling $q$ of the scalar field. At early times, simulations reveal
that for sufficiently small $q\lesssim \frac{3r_H}{QL}$, the field
decays, resulting in rapid ringdown to RN-AdS. For
slightly larger $q$, a growing mode is present, and the instability
ensues. We checked that the initial growth rate---while the
perturbation remains small---matches the prediction
of~\cite{2011PhRvD..83f4020U} in the linearized case. At later times,
the perturbation becomes nonlinear, as backreaction on the BH
becomes significant, with the spacetime eventually settling into a
stationary hairy BH.

The superradiant cases display the following characteristics: (i) The
scalar field eventually saturates in amplitude and has harmonic time
dependence\footnote{Previously constructed hairy black holes
  \cite{Basu:2010uz,Dias:2011tj} have static $\psi$. This difference
  arises because these works make the gauge choice that the Maxwell
  field vanishes at the horizon, whereas we set it to zero at
  infinity.}, resulting in a \emph{time-independent} stress-energy
tensor and metric; (ii) Significant amounts of charge [measured at
the apparent horizon (AH)] are extracted from the BH by the scalar
field, with more extracted at larger $q$; (iii) The irreducible mass
of the BH (proportional to the square root of the area)\footnote{As
  the stationary stage is reached, the AH location coincides with the
  event horizon.}  approaches that of a Schwarzschild-AdS BH of mass
$M$, with closer approach for larger $q$ (implying less mass
extraction for larger $q$); (iv) The scalar hair is distributed
farther away\footnote{This observation highlights the difference
  between placing an artificial boundary at some location with respect
  to the BH versus at the boundary of AdS.} from the BH for larger
$q$; and (v) The approach to the final state is less smooth for larger
$q$, in a sense that will be described below.

\begin{figure}[tb]
  \centering
    \includegraphics[width=4cm,height=3.6cm]{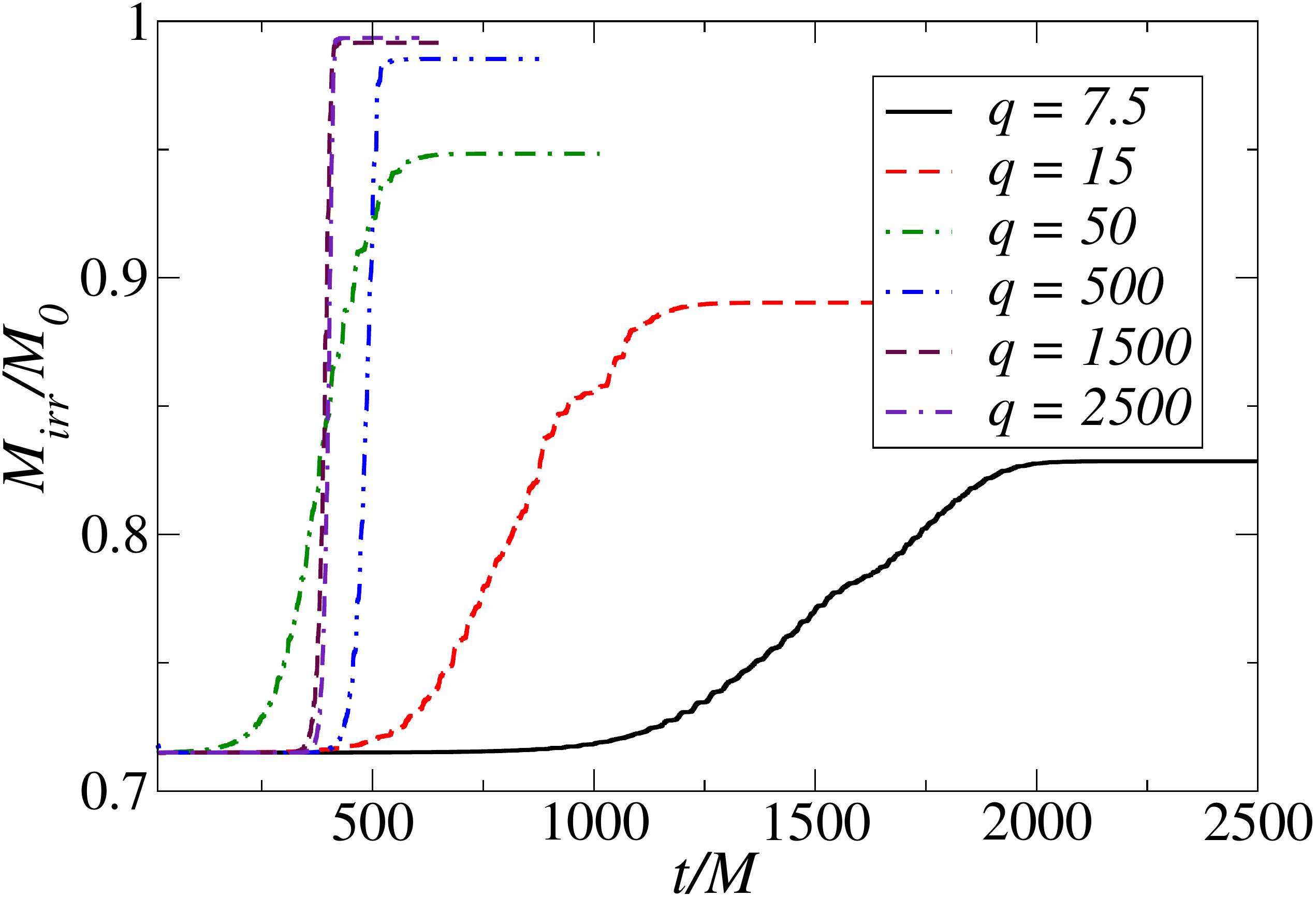}
    \includegraphics[width=4cm,height=3.6cm]{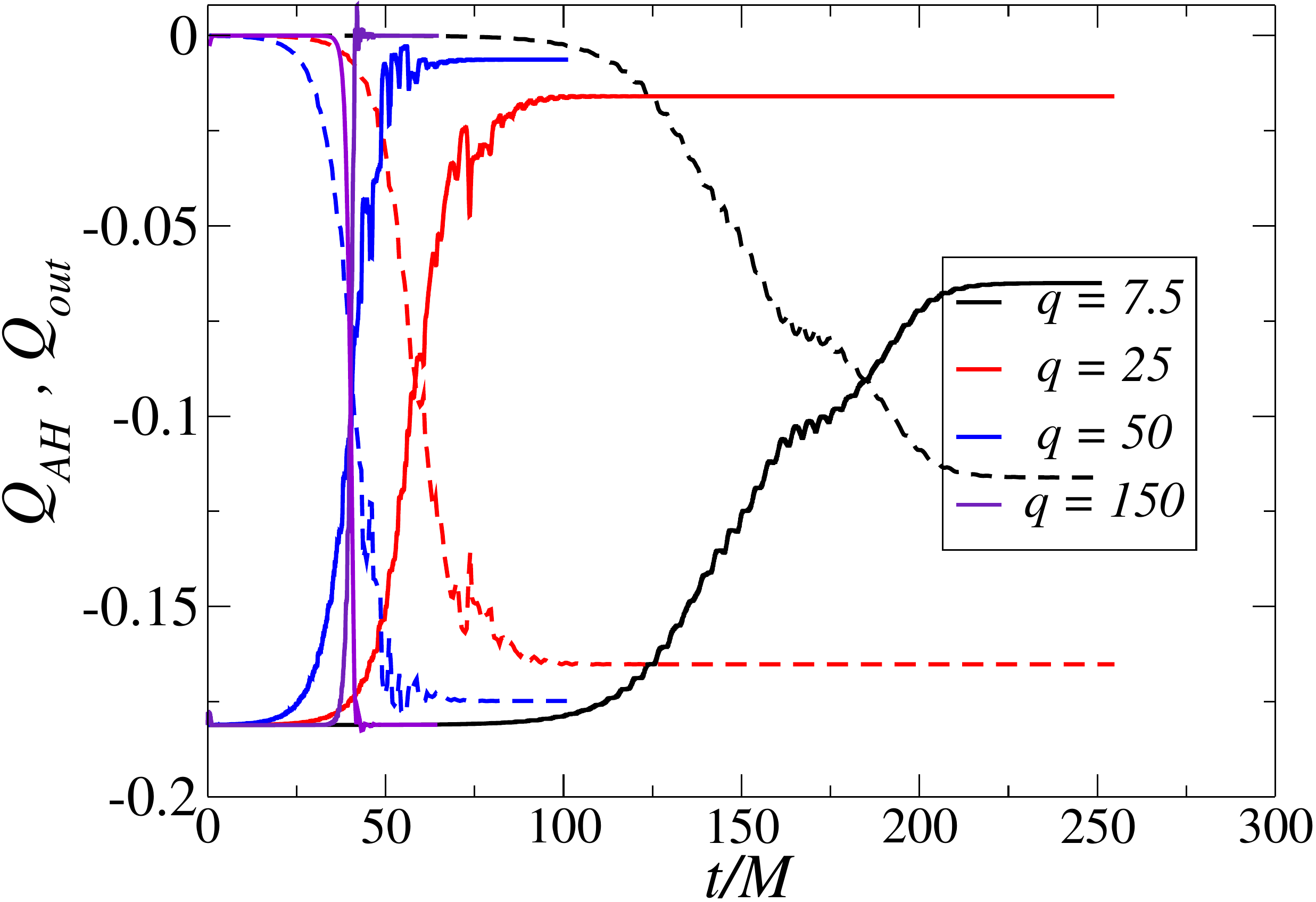}
    \caption{\label{mqj_time}(Left)(Normalized) Irreducible mass
      versus time for representative values of $q$. As $q$ is
      increased the growth rate increases as well.  (Right) Charge
      within the AH (solid), and the charge of the
      scalar field outside on a constant-$v$ slice (dashed), vs time. 
      As the value of $q$ is increased, the
      dynamical time scale shortens. For even smaller $q$ (not shown),
      curves are very smooth, with no steps or oscillations. (Note
      that for the initial data employed, for the largest-$q$ case,
      most of the scalar field energy falls immediately into the BH
      as it backscatters off itself, resulting in a smaller
      effective initial perturbation.)}
\end{figure}

\begin{figure}[tb]
  \centering
    \includegraphics[width=4cm,height=4cm]{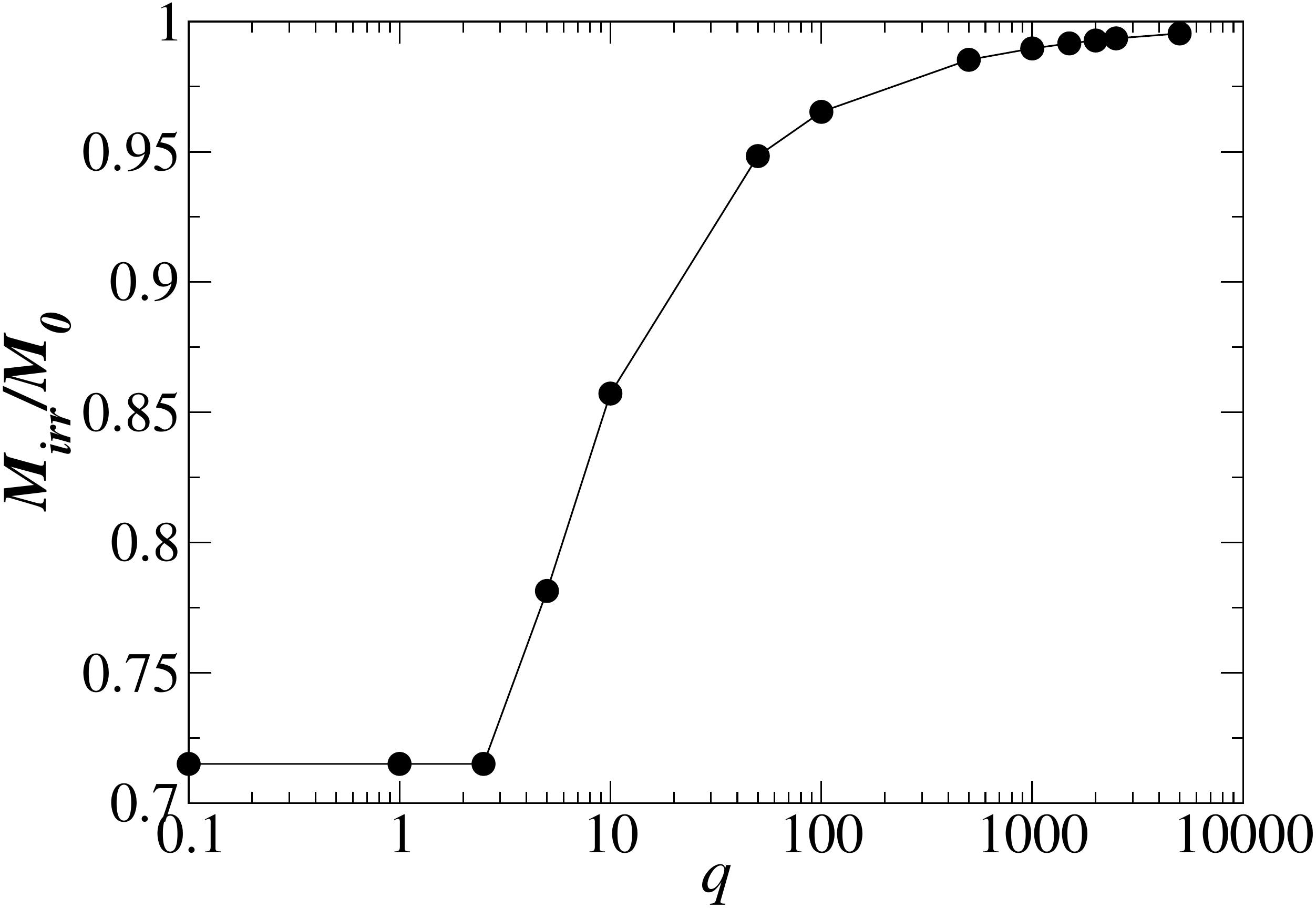}
    \caption{\label{m_irreducible}Normalized irreducible mass vs
      $q$.  As $q$ is increased, $M_{\rm irr}/M_0 \rightarrow 1$.
      This is consistent with the claim that the final BH state for
      sufficiently large values of $q$ is a Schwarzschild-AdS black
      hole of mass $M$ (the ADM mass), surrounded by a distant high
      charge, low-mass scalar field condensate.}
\end{figure}

In Figure~\ref{mqj_time}, we show the irreducible mass and charge of
the BH vs.~time, for various choices of $q$. In
the left figure, we compare the irreducible mass $M_{\text{irr}}$ of
the AH of our dynamical BH with that of a
(uncharged) Schwarzschild-AdS BH with mass $M$ (we denote this
quantity $M_0$). As expected from the area theorem, the irreducible
mass (entropy) never decreases. For small $q$, $M_{\text{irr}}$ displays very
smooth approach to its final value, while for large $q$, it displays
some step-like behavior prior to reaching a plateau. Moreover, for large $q$, the ratio
$M_{\text{irr}}/M_0\to1$, as depicted in Figure~\ref{m_irreducible}.

The right side of Figure~\ref{mqj_time} shows the charge, both of the
AH, and the integrated charge (at each constant time surface and outside the AH) of the scalar field (by
charge conservation, these must sum to the ADM charge $Q$). The
charge displays much more interesting behavior than the irreducible mass,
including some up-and-down oscillations. As with $M_{\text{irr}}$, the
low-$q$ case is smoothest. We find that for large $q$, nearly all of
the charge can be extracted from the BH. This contrast between
the mass and charge extracted arises because the charge-to-mass ratio
of the final state scalar field mode is larger for larger $q$. 

\begin{figure}[tb]
  \centering
    \includegraphics[width=5cm]{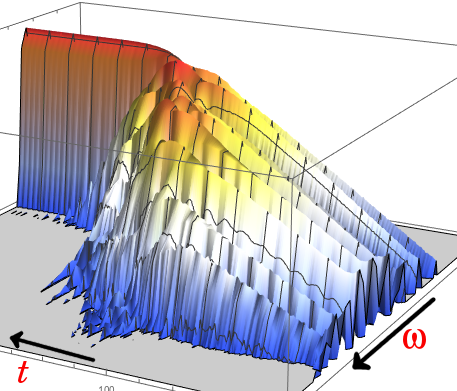}
    \caption{Spectrogram showing logarithm of
      amplitude of Fourier transform of $\varphi_3(v)$ as a function
      of time. This is computed by partitioning the time axis into
      intervals of length $\Delta v = 4\pi$, and performing discrete
      Fourier transforms on these intervals. The intervals
      overlap, with starting points offset by $\delta v=\pi/8$. The case
      here has $q=12$, $r_H=0.2$, and $Q$ set to $80\%$ of the
      extremal value.  At early times, the lowest eight modes grow
      exponentially, with faster growth for lower frequencies.  As
      charge and mass is extracted, all modes (aside from the
      fundamental) eventually start to decay, with higher frequencies decaying
      first. The growth rate of the fundamental approaches zero,
      leaving a final static BH with a harmonically oscillating scalar
      condensate.\label{spectra}}
\end{figure}

The less smooth behavior of the instability at large $q$ can be
understood by studying the mode content. In Figure~\ref{spectra} we
show a spectrogram of an evolution, where we plot the frequency
content of $\varphi_3(v)$ vs time. This reveals the individual modes
present in the solution, and their growth or decay as a function of
time. The initial data contain a large number of modes, with real
frequencies $\Re(\omega_n) \approx (2n+3)/L + C$ in the small BH
limit. Most of these do not satisfy the superradiance condition, and
they decay rapidly~\cite{2011PhRvD..83f4020U}. The lower frequency
modes, however, with $(2n+3)\lesssim \frac{qQL}{r_H}$, are unstable,
and are visibly growing exponentially in the spectrogram
[lower-$\Re(\omega_n)$ modes grow more rapidly]. (In
Figure~\ref{spectra} there are eight such modes.)  Over time, these modes
extract charge and mass from the BH, and one by one (starting at large
$n$) they begin to decay, and are re-absorbed by the BH. In the end,
the BH is discharged to the level where the fundamental ($n=0$) mode
has zero growth rate, and it remains as the condensate.

The oscillations of Figure~\ref{mqj_time} can be now understood as an
effect of having a mixture of modes, some of which are extracting, and
others depositing, charge and mass into the BH. In the end, the BH
reaches the hairy state, with all the higher modes having
decayed. Similar oscillations for larger-$q$ evolutions were
misinterpreted as a ``bosenova'', or explosion,
in~\cite{Sanchis-Gual:2015lje}. (The bosenova of
\cite{2011PhRvD..83d4026A,Yoshino:2012kn} arises instead when axion
self-interactions cause a collapse of the axion field.) Indeed, for
$q$ just above the instability threshold, there is only a single mode,
and evolution displays a very smooth approach to the stationary end
state.

\begin{figure}[tb]
  \centering
    \includegraphics[width=4cm,height=3.6cm]{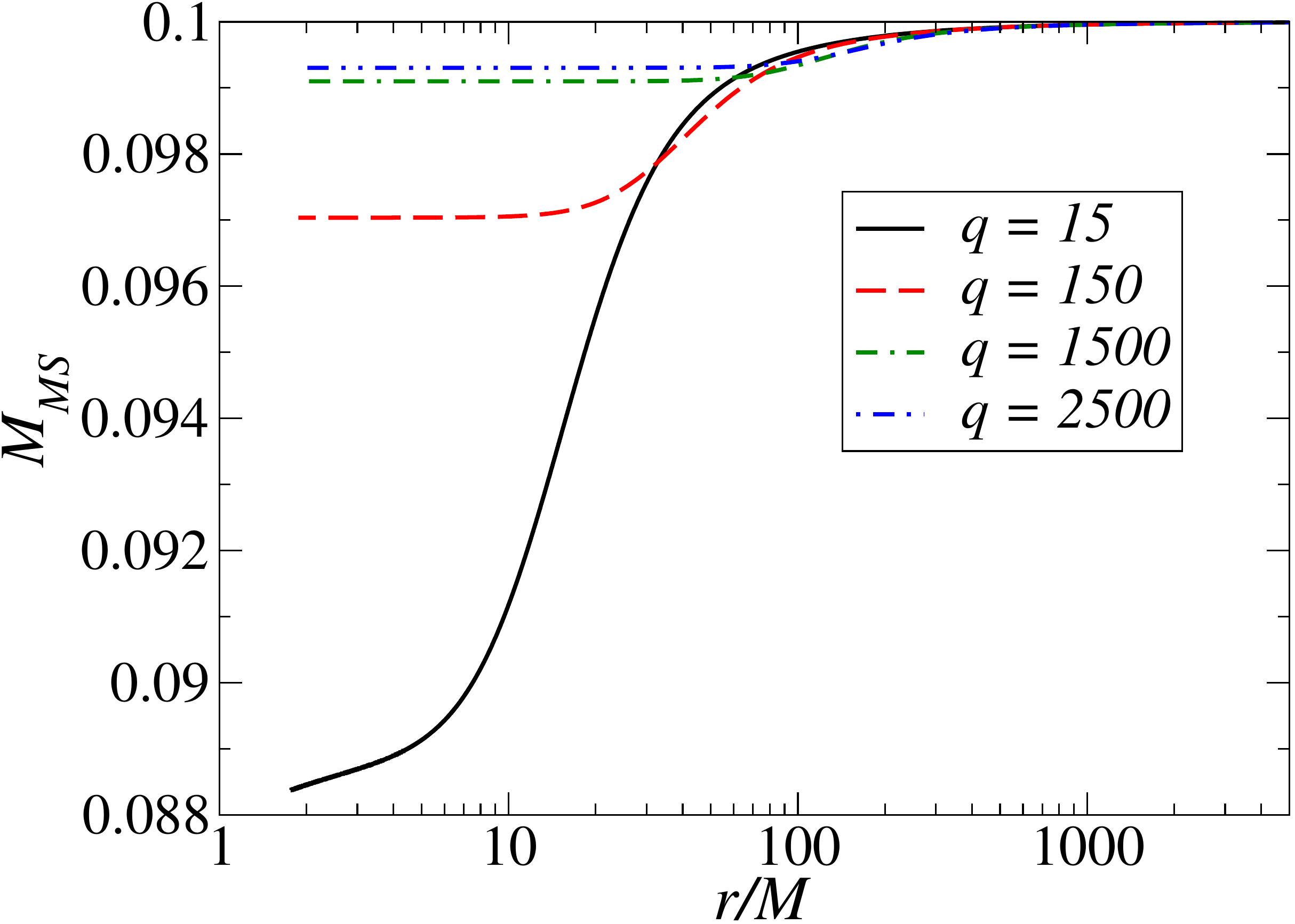}
    \includegraphics[width=4cm,height=3.6cm]{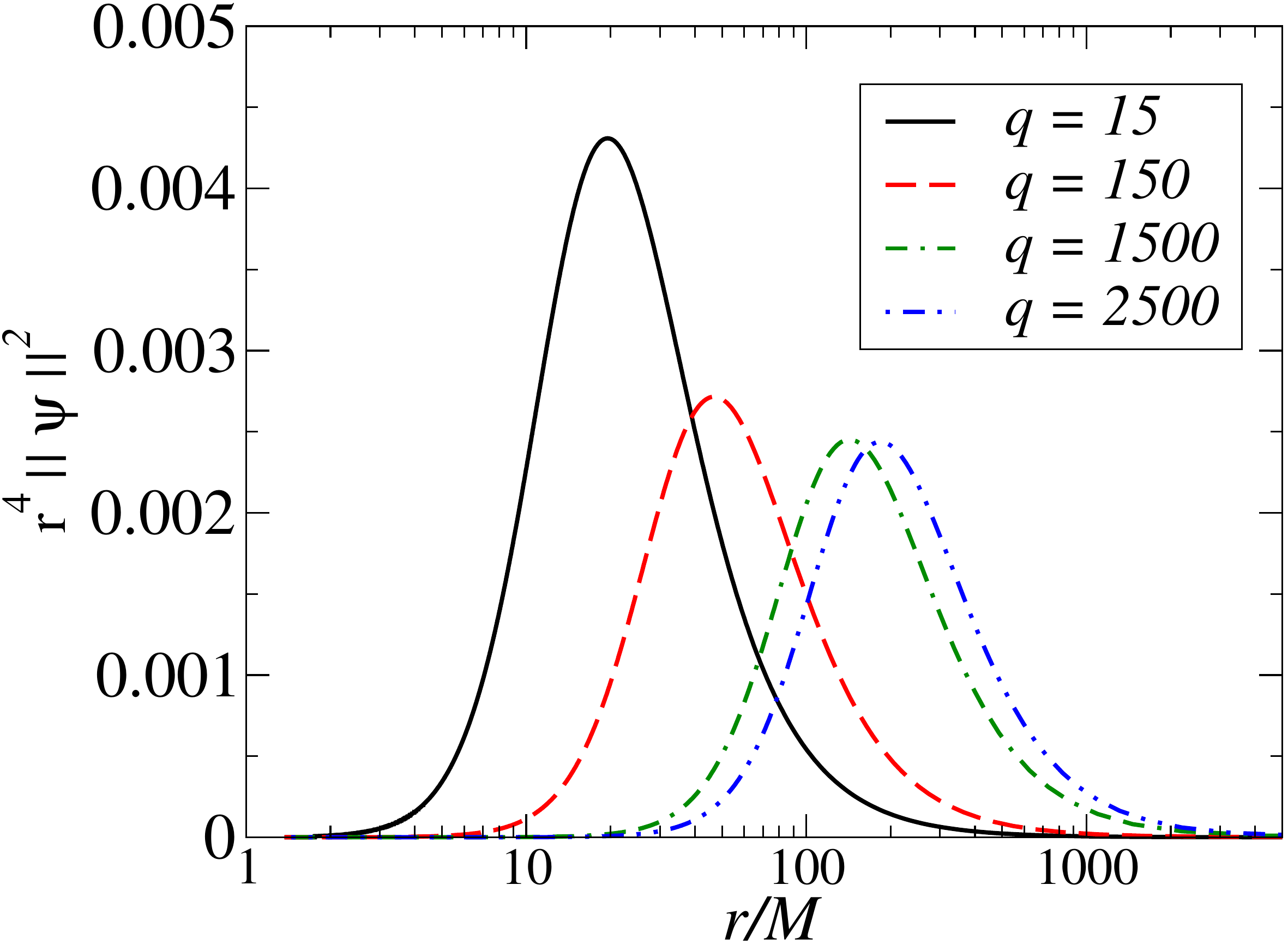}
    \caption{(Left)Misner-Sharp mass versus radius for
      representative values of $q$, measured for the late-time static
      spacetime. For large values of $q$, $M_{\rm MS}$ is constant to
      relatively large radial distance, indicating that BH is
      essentially uncharged \emph{and} the scalar field hair lies far
      away from it.  At larger radii, a radial dependence
      arises because of the presence of both the electromagnetic and
      scalar fields.  (Right) Norm (squared and rescaled by $r^4$)
      of the scalar field, also at late times. The field is localized
      far from the BH for large values of $q$.\label{misner}}
\end{figure}

Radial profiles of the final state are illustrated in
Figure~\ref{misner}. On the left, we plot the Misner-Sharp mass
$M_{\text{MS}}(r)$ (suitably defined so as to take into account the
contribution of the AdS curvature~\cite{Maeda:2012fr}) as a function of
radius. As $q$ is increased, $M_{\text{MS}}$ increases in value at the
AH, again confirming that less mass is extracted. As 
$\mathscr{I}$ is approached, $M_{\text{MS}}\to M$. Moreover, for large $q$,
$M_{\text{MS}}$ is constant in $r$ near the BH, while for small $q$ it
grows. This, together with the right figure, shows that the scalar
field condensate and the electric field are localized farther away
from the BH for larger $q$.

Finally, Figure~\ref{m_irreducible} shows the normalized irreducible
mass of the final state BH as a function of $q$. As $q$ is increased,
$M_{\text{irr}}/M_0\to1$. (For $q=5000$,
$M_{\text{irr}}/M_0\approx99.5\%$.) This, together with the radial
profile information, indicates that at late times and large $q$, the
BH region approaches Schwarzschild-AdS, surrounded by a distant
low-mass high-charge condensate.

{\bf Final words.}---We have described the full dynamical behavior of
the charged superradiant instability in AdS. Initially, superradiant
modes extract charge and mass from the BH and grow
exponentially~\cite{2015LNP...906.....B}. As this process unfolds, the
higher-frequency modes cease to be superradiant, and fall back into
the BH, returning energy and charge and resulting in nontrivial
dynamics. Eventually, the fundamental mode remains as a condensate,
with zero growth. While nonlinear couplings between modes (via gravity
and electromagnetism) can generate higher-frequency modes in
AdS~\cite{Bizon:2011gg,Balasubramanian:2014cja}, any such processes
are overwhelmed by the fact that in the end, all modes beyond the
fundamental are decaying. The ultimate fate is a stable hairy BH, with
the scalar condensate distributed far away from the BH for large $q$.

Among the scalar modes, the final mode that remains maximizes the
charge-to-mass ratio $q/\omega$, and its growth corresponds to
maximizing the entropy increase of the BH. This observation may
elucidate possible behavior of the superradiant instability in the
rotating case. For a Kerr-AdS BH, the stability criterion for a mode
is $\omega < m\Omega_H$, where $\Omega_H$ is the angular frequency of
the BH, and $m$ the azimuthal number of the
perturbation~\cite{Hawking:1999dp}. This is comparable with the
condition $\omega r_H < qQ$ in the charged case, but with the key
difference that $m$ can take any integer value, whereas $q$ is a fixed
parameter of the system. In the rotating case, a given perturbation
may be expected to spin down the black hole to the point where only
the most superradiant mode remains (i.e., that which causes the black
hole to maximize its entropy), but now just marginally stable. It
would be interesting to examine the superradiant mode frequencies in
Kerr-AdS presented in~\cite{Cardoso:2013pza} from this point of view;
it is plausible that the final (i.e., most superradiant) mode has
$m\to\infty$, consistent with speculation of~\cite{Niehoff:2015oga}.

Finally, in astrophysical applications, the outer potential barrier is
no longer infinite (as in AdS), and is instead typically provided by a
mass term for the
field~\cite{Damour:1976kh,Detweiler:1980uk,2011PhRvD..83d4026A}. Such
a case provides a cutoff in mode energy, and on the efficiency of
energy extraction.

\medskip {\bf Acknowledgments.}---We would like to thank A.~Buchel,
S.~L.~Liebling, O.~Sarbach and J.~Winicour for discussions and
comments throughout this project.  This work was supported in part by
NSERC through a Discovery Grant (to L.L.), by CIFAR (to L.L.), by
CONACyT-Mexico (to P.B.), and by Perimeter Institute for Theoretical
Physics. Research at Perimeter Institute is supported by the
Government of Canada through Industry Canada and by the Province of
Ontario through the Ministry of Research and Innovation.

\newpage

\onecolumngrid  \vspace{1cm} 
\begin{center}  
{\Large\bf Supplemental Material} 
\end{center} 
\appendix 
\tableofcontents  

\setcounter{section}{19}
\setcounter{equation}{0}
\section{Numerical method}

Here we provide additional details of our numerical approach. Our
implementation is similar to that of~\cite{Chesler:2013lia}, but with several
modifications: (1) We include Maxwell and complex scalar fields, (2)
Our black hole has spherical (as opposed to planar) topology, (3) We
excise the black hole from the computational domain and
(4) instead of pseudospectral methods we employ finite differences with derivatives
satisfying summation by parts~\cite{Calabrese:2003yd,Buchel:2012gw}.

As noted in the main text, in order to discretize the spacetime, we
adopt a compact spatial domain so as to include the AdS boundary ($\mathscr{I}$) in our computational
grid and ensure the full equations are consistently implemented. Hence, we introduce
$\rho=1/r$ as our spatial coordinate and define new variables,
\begin{align}
  A(v,r) &= \alpha(v,1/r) = \alpha(v,\rho),\\
  \Sigma(v,r) &= \sigma(v,1/r) = \sigma(v,\rho),\\
  d_+\Sigma(v,r) &= s(v,1/r) = s(v,\rho),\\
  W(v,r) &= \nu(v,1/r) = \nu(v,\rho),\\
  \psi(v,r) &= \varphi(v,1/r) = \varphi(v,\rho),\\
  d_+\psi(v,r) &= \Pi(v,1/r) = \Pi(v,\rho).
\end{align}
From the asymptotic forms \eqref{eq:Aasymp}--\eqref{eq:psiasymp}, we
see that some of these fields diverge at $\mathscr{I}$, so it is more
convenient to evolve ``hatted'' fields that have better asymptotic
behavior. We define
\begin{align}
  \hat{\alpha} &= \alpha - \frac{1}{L^2}\sigma^2 - 1,\\
  \hat{\sigma} &= \sigma - \frac{1}{\rho},\\
  \hat{s} &= s - \frac{1}{2L^2}\sigma^2-\frac{1}{2},\\
  \hat{\nu} &= \nu,\\
  \hat{\varphi} &= \frac{1}{\rho^2}\varphi,\\
  \hat{\Pi} &= \frac{1}{\rho}\Pi.
\end{align}
In order to integrate a first order system [\eqref{eq:Einstein2} and
\eqref{eq:Einstein4} have second order space derivatives], we also
introduce $\hat{\beta} = \partial_\rho\hat{\alpha}$ and
$\hat{z} = \partial_\rho\hat{w}$. With the compact domain, the
asymptotic conditions are simply imposed as boundary conditions at $\rho=0$
(see Table~\ref{table:bc}).

\begin{table}[t]
  \begin{ruledtabular}
    \begin{tabular}{ c c c}
      $f$ & $f|_{\rho=0}$ & $\partial_\rho f|_{\rho=0}$ \\
      \hline
      $\hat{\alpha}$ & $-L^2\dot\lambda$ & $-2M$ \\
      $\hat{\sigma}$ & $\frac{1}{2}L^2\lambda$ & $0$ \\
      $\hat s$ & $0$ & $-M$ \\
      $\hat w$ & $\nu$ & $Q$ \\
      $\hat \varphi$ & $0$ & $\varphi_3$ \\
      $\hat \Pi$ & $0$ & $-\frac{3}{2L^2}\varphi_3$\\
      $\hat \beta$ & $-2M$ & $2L^2\lambda M + \frac{Q^2}{2}$\\
      $\hat z$ & $Q$ & $-L^2\lambda Q$\\
    \end{tabular}
  \end{ruledtabular}
  \caption{\label{table:bc}Boundary conditions for the evolved fields.}
\end{table}

The equations of motion for the hatted fields are obtained by
substitution into the equations already written in the main text. We
omit the resulting (lengthy) expressions. Given $\hat\varphi$ at time
$v=v_0$ (more specifically, we keep track of its real and imaginary
parts), we integrate eqs.~\eqref{eq:Einstein1}, \eqref{eq:Einstein2},
\eqref{eq:Einstein4}, \eqref{eq:Maxwell2}, and \eqref{eq:Scalar1},
together with $\hat{\beta} = \partial_\rho\hat{\alpha}$ and
$\hat{z} = \partial_\rho\hat{w}$, to obtain all of the remaining
hatted fields at time $v=v_0$. These equations are all first order in
space, so we impose the initial value for each field at $\rho=0$, as
listed in the middle column of Table~\ref{table:bc}. To evolve forward
in time, we make use of the definition
\begin{equation}
  d_+\psi = \partial_v\psi + \frac{1}{2}A\partial_r\psi
\end{equation}
(again, re-written in terms of hatted fields), to obtain
$\partial_v\hat\varphi$ at time $v=v_0$. Knowing this, we evolve
forward one step in time, and repeat the procedure.

The boundary conditions in Table~\ref{table:bc} contain two free
functions, $\lambda$ and $\nu$, which correspond to the remaining
freedom in the choice of radial coordinate and the vector potential
after imposing \eqref{eq:metricform} and \eqref{eq:Maxwellform},
respectively. We set these quantities to zero, so the only remaining
boundary conditions are the (constant) ADM mass and charge ($Q$ and
$M$), which we are free to choose. The final function, $\varphi_3(v)$,
shown in the table, is not set in advance, but rather is an
\emph{output} of the evolution, which may be read off.

The full equations are discretized by adopting a uniform grid
$\rho_i = (i-1) d\rho$ with $d\rho = \rho_{\rm inner}/(N-1)$ (with
$\rho_{\rm inner} < \rho_{\rm AH}$ the location of the AH; see
below). Null hypersurfaces are discretized in time separated by
$dv = \gamma d\rho$ with $\gamma$ chosen typically $=0.4$ to ensure
satisfying the CFL condition.  Radial integrations are performed using
a 4th order Runge-Kutta, being mindful of a few subtleties. Some of
the equations of motion involve $0/0$-type terms at $\rho=0$, in these
cases we impose the radial derivative explicitly at the boundary (and
where needed regularize the equations via the L'H\^opital rule). (For
this reason, we provide the third column of Table~\ref{table:bc}.)
Additionally, the equation of motion for $\hat s$ is of the form
$\partial_\rho \hat s \sim \hat s/\rho + \ldots$, which is problematic
as its regularization leaves no equation. We instead use a 2nd order
implicit scheme for this equation, making the spatial integration 2nd
order as a whole. Time integrations are performed with RK4, and
spatial derivatives satisfy the summation by parts property and are
2nd order accurate at boundary points while 4th order accurate at
interior ones~\cite{Calabrese:2003yd}.

As mentioned, we do not exploit the freedom in $\lambda$ to fix the
position of the AH at a constant radial coordinate as
in~\cite{Chesler:2013lia}. Rather, we find the AH at each incoming
null surface and excising a number of grids points inside of it
(typically $10$ points; see, e.g.,~\cite{Buchel:2012gw,Gomez:1997pd}).

\section{Code validation}

We have confirmed the validity of our implementation through extensive
tests summarized below:
\begin{itemize}
\item By choosing small values of $q$ and a configuration for the
  scalar field not satisfying the superradiance condition we have
  confirmed the decay of the perturbations and with a rate in
  agreement with those presented in~\cite{2011PhRvD..83f4020U}. Our
  extracted values for both real and imaginary parts of $\omega$
  agrees with those obtained in linearized perturbations to better
  than $0.01\%$ even with relatively modest resolutions ($N=801$).
\item Choosing instead larger values of $q$ we obtained the onset of
  the superradiance instability and confirmed our extracted values
  agree with those in~\cite{2011PhRvD..83f4020U} with high accuracy.
\item An even more stringent test is provided by evaluating
  independent residuals---furnished by the equations
  \eqref{eq:Einstein3} and \eqref{eq:Maxwell1}. That is, we
  independently evaluate these equations, discretized in a
  straightforward second order approximation, employing fields at two
  sequential time slices, and confimed the equations are satisfied to
  this order throughout our evolutions.
\item All results presented here have been validated by employing
  several distinct resolutions. Typical resolutions of $N=1601$ points
  suffice for excellent accuracy for low to mid values of $q$ while we
  adopt higher ones $N=3201$ or $N=4801$ for $q>1000$.
\item We performed self-convergence tests showing that our numerical solutions
converge at least to second order.
\end{itemize}

\bibliography{mybib}

\end{document}